\documentclass[aps,prl,twocolumn,superscriptaddress]{revtex4}

\usepackage{graphicx}
\begin{document}

%\preprint{}

\title{Doppler cooling of calcium ions using a dipole-forbidden transition}


\author{Richard J. Hendricks}
\author{Jens L. S\o rensen}
\affiliation{QUANTOP --- Danish National Research Foundation Center for Quantum Optics, Department of Physics and Astronomy, University of Aarhus, DK-8000 \AA rhus C., Denmark}
\author{Caroline Champenois}
\author{Martina Knoop}
\affiliation{Physique des Interactions Ioniques et Mol\'{e}culaires (CNRS UMR 6633),
Universit\'{e} de Provence, Centre de Saint J\'{e}r\^{o}me, Case C21, 13397 Marseille Cedex 20, France}
\author{Michael Drewsen}
\email[]{drewsen@phys.au.dk}
\affiliation{QUANTOP --- Danish National Research Foundation Center for Quantum Optics, Department of Physics and Astronomy, University of Aarhus, DK-8000 \AA rhus C., Denmark}

%\thanks{}

\date{\today}

\begin{abstract}
Doppler cooling of calcium ions has been experimentally demonstrated using the S$_{1/2}$$\longrightarrow$D$_{5/2}$ dipole-forbidden transition. Scattering forces and fluorescence levels a factor of 5 smaller than for usual Doppler cooling on the dipole allowed S$_{1/2}$$\longrightarrow$P$_{1/2}$ transition have been achieved. Since the light scattered from the ions can be monitored at (violet) wavelengths that are very different from the excitation wavelengths, single ions can be detected with an essentially zero background level. This, as well as other features of the cooling scheme, can be extremely valuable for ion trap based quantum information processing.

\end{abstract}

%\keywords{}

\maketitle

%%%%%%%%%%%%%%%%%%%%

In the last few decades, a large variety of laser cooling techniques have been proposed and demonstrated~\cite{Adams:1997,Leibfried:2003}.  These include Doppler cooling~\cite{Prodan:1982,Chu:1985}, sideband cooling~\cite{Neuhauser:1978,Diedrich:1989,Monroe:1995,Roos:2000}, stimulated cooling~\cite{Aspect:1986}, polarization gradient cooling~\cite{Lett:1988,Dalibard:1989,Ungar:1989}, magnetically-induced laser cooling~\cite{Shang:1990}, Raman cooling~\cite{Kasevich:1992,Reichel:1995} and intercombination line cooling~\cite{Binnewies:2001,Curtis:2001}. A characteristic of these cooling schemes is a trade-off between the velocity range where the cooling mechanism is efficient and the final achievable minimum temperature. As a consequence, most experiments today use more than one cooling mechanism. For quasi-free neutral atoms with degenerate ground state sublevels, a combination of Doppler and polarization gradient cooling can relatively easily be used to obtain dense ensembles of atoms in the micro-kelvin range~\cite{Lett:1988,Salomon:1990}. For neutral atoms lacking ground state sublevels, such as most alkali earth elements, a more elaborate cooling scheme such as a second Doppler cooling stage on a weak intercombination line is required in order to obtain ultra cold atoms~\cite{Binnewies:2001,Curtis:2001,Mukaiyama:2003}. With respect to trapped ions, a two-step cooling process consisting of a Doppler cooling phase followed by a resolved-sideband cooling phase is typically used to cool from initial high thermal energies to close to the ground state of the trapping potential.  For the sideband cooling step either a Raman cooling scheme~\cite{Monroe:1995}, an EIT cooling process~\cite{Roos:2000} or excitation on a weak, narrow-linewidth transition~\cite{Diedrich:1989,Roos:1999} is used. 

\begin{figure}
\includegraphics[width=0.48\textwidth]{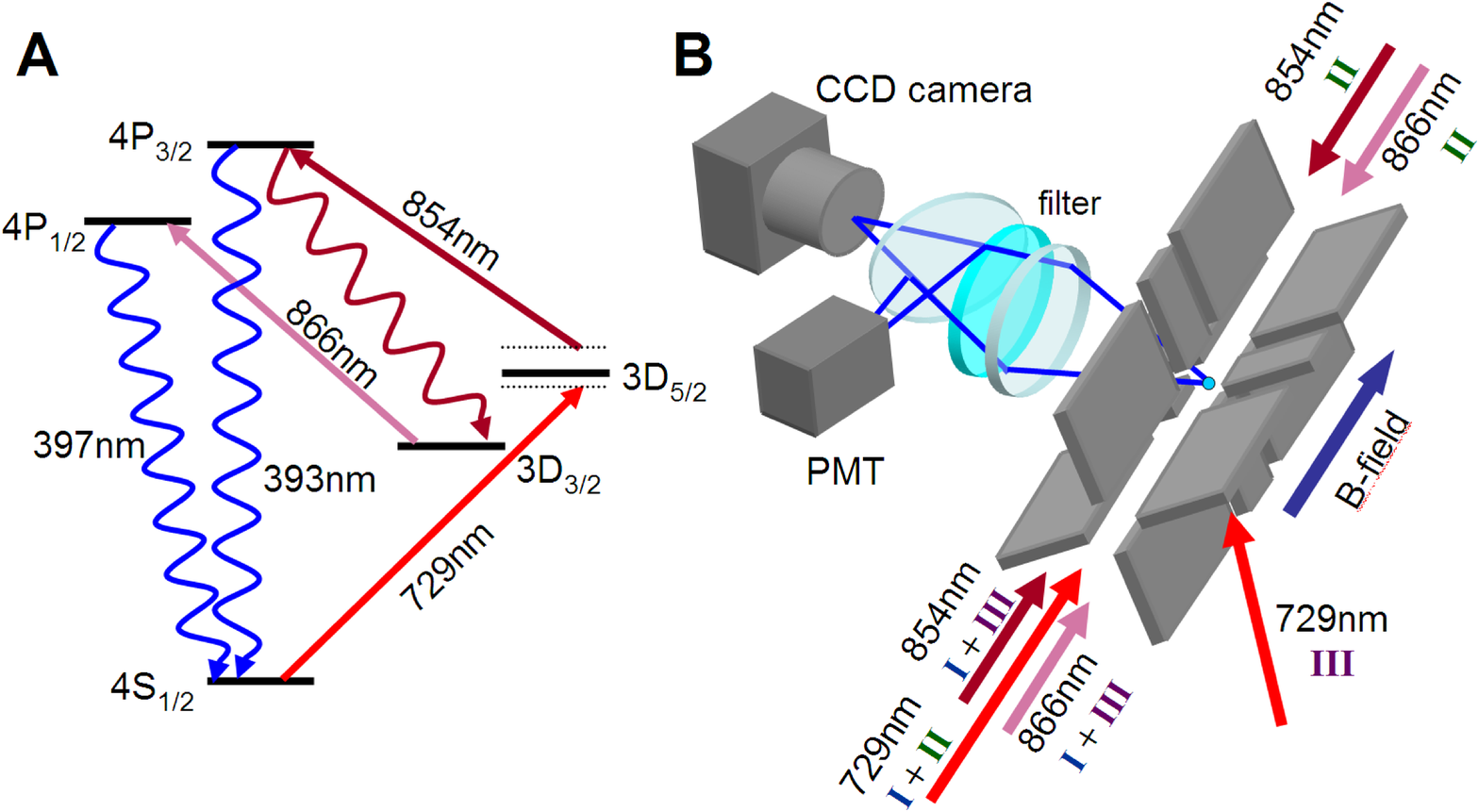}
\caption{\label{fig:setup}(Color online.) \textbf{A}~Relevant energy levels and transitions of the ${}^{40}$Ca$^+$ ion.  \textbf{B}~Experimental setup.  Three different situations are illustrated: I.~Axial cooling where the cooling and assisting lasers are co-propagating.  II.~Axial cooling with counter-propagating cooling and assisting lasers.  III.~3-dimensional cooling with the cooling laser at an angle of 45$^\circ$ to the trap axis and the assisting laser parallel to the trap axis.}
\end{figure}

In this Letter, we demonstrate a scheme for Doppler cooling ${}^{40}$Ca$^+$ ions that relies on the combined action of driving the S$_{1/2}$$\longrightarrow $D$_{5/2}$ electric quadrupole transition and the dipole-allowed D$_{5/2}$$\longrightarrow$P$_{3/2}$  transition (figure~\ref{fig:setup}A).  While the required velocity dependence of the cooling force is obtained through excitation of the S$_{1/2}$$\longrightarrow $D$_{5/2}$ quadrupole transition by an intense laser beam at 729~nm, a co-propagating 854~nm `assisting' laser beam is applied to drive the D$_{5/2}$$\longrightarrow$P$_{3/2}$ dipole transition.  This latter laser field increases the effective decay rate $\Gamma^{\prime}$ of the D$_{5/2}$ state to the S$_{1/2}$ ground state, as well as increasing the cooling force because of the extra momentum transferred to the ions by the 854~nm photons. Ground state cooling using a resolved sideband of the S$_{1/2}$$\longrightarrow $D$_{5/2}$ quadrupole transition has already been demonstrated in the regime where $\Gamma^{\prime}$ is smaller than the relevant motional mode frequency of the trapped ions ~$\omega$~\cite{Roos:1999}.  In contrast to this, we here consider Doppler cooling, where $\Gamma^{\prime} > \omega$. By using a narrow-bandwidth 729~nm laser, lowering the intensities, and changing the detunings of the lasers the presented cooling scheme can, however, be transformed into a resolved-sideband cooling scheme with little additional complexity.  In our experiments, scattering forces and fluorescence levels only about a factor of 5 smaller than for usual Doppler cooling on the dipole-allowed S$_{1/2}$$\longrightarrow$P$_{1/2}$ transition have been achieved. As discussed at the end of this Letter, the Doppler cooling scheme presented here has several features of particular interest for quantum information processing with trapped ions.
%%%%%%%%%%%%%%%%%%%%

A reduced level scheme of Ca$^+$ is shown in figure~\ref{fig:setup}A, together with the laser excitations and fluorescence paths of interest for the presented cooling scheme. The key laser-excited transitions in the cooling scheme are the S$_{1/2}$$\longrightarrow$D$_{5/2}$ electric quadrupole transition at 729~nm, which has a natural width of just $\Gamma_c = 2\pi \times 0.14$~Hz~\cite{Barton:2000}, and the 854~nm D$_{5/2}$$\longrightarrow$P$_{3/2}$ dipole transition, which has a linewidth of $\Gamma = 2\pi\times 23$~MHz.  Ions excited to the P$_{3/2}$ level will predominantly decay to the S$_{1/2}$ ground state by the emission of light at 393~nm.  They can also, however, decay with probabilities 0.07 and 0.008 to the D$_{5/2}$ and D$_{3/2}$ levels, respectively, emitting infra-red light. To prevent optical pumping into the D$_{3/2}$ level, a repumper laser at 866~nm is used to excite ions to the P$_{1/2}$ level, from where the ground state can be reached via the emission of a 397~nm photon. 

%%%%%
%%%%%

The experimental setup is sketched in figure~\ref{fig:setup}B.  The $^{40}$Ca$^+$ ions are stored in a segmented linear radio-frequency trap (details of which can be found in~\cite{Sorensen:2006}) housed in an ultra-high vacuum chamber with a residual pressure of approximately $6$$\times$$10^{-10}$~mbar. The ions are loaded into the trap via isotope-selective photo-ionization~\cite{Kjaergaard:2000}, and the trap parameters are set such that the single ion axial and radial motional frequencies are approximately $2\pi \times 0.56$~MHz and $2\pi \times 0.95$~MHz, respectively. Sets of Helmholtz coils are used to null the magnetic field to a level of about 5~mG.  

The 729~nm light, which has a linewidth of approximately 200~kHz, is generated by a Ti:Sapphire laser stabilized to an external optical cavity. Up to 250~mW of power is available in a beam focused to a waist of 50~$\mu$m at the trap center. This leads to a maximum achievable quadrupole transition Rabi frequency of $\sim$$2\pi \times 1$~MHz~\cite{James:1998}. The beam propagates either parallel to or at an angle of 45$^\circ$ to the axis of the trap, with the latter configuration enabling cooling of both the axial and radial motion of the trapped ions. The light is always linearly polarized within the horizontal plane. The 854~nm assisting and the 866~nm repumper laser beams are derived from extended-cavity diode lasers locked to external optical cavities and have linewidths of about 300~kHz.  A maximum of 3~mW is available at each wavelength, and beam waists of $\sim$280~$\mu$m are used. Both beams can either co- or counter-propagate with the 729~nm laser beam when the latter is directed along the axis of the trap.  The polarization of both these lasers is rotated at a frequency of 4~MHz using an electro-optic modulator to prevent the formation of `dark' states~\cite{Barwood:1998}.  In order to keep the AC Stark shift and broadening of the D$_{5/2}$ level Doppler insensitive at low ion velocities, the 854~nm laser is kept at a constant detuning of approximately 100~MHz.  With a power of 1~mW this leads to an effective decay rate of the D$_{5/2}$ level $\Gamma^{\prime}$ of about $2\pi \times 2$~MHz, which is larger than the single ion oscillation frequencies.  

Since the laser power at 729~nm is limited and the overlap between the trapping volume and the 729~nm laser beam is small, cooling of the ions using the 729~nm and 854~nm lasers is not very efficient in the phase immediately after loading where the ions have kinetic energies in the eV range. For this reason, in all experiments the ions are first Doppler cooled in all directions using the S$_{1/2}$$\longrightarrow$P$_{1/2}$ dipole-allowed transition at 397~nm in the presence of the 866~nm repumper. Mechanical shutters with a switching time of approximately 0.1~ms are used to change between the 397~nm and 729~nm cooling configurations.

Ions are detected by imaging the 393~nm and 397~nm fluorescence light onto both a photo-multiplier tube (PMT) based photon-counting module and an image intensified \textrm{CCD} camera.  Filters are used to prevent any light at red or infrared wavelengths from reaching the detectors. The overall detection efficiencies of the photon-counting module and the CCD camera system are $3.6$$\times$$10^{-4}$ and $8.4$$\times$$10^{-5}$, respectively.

\begin{figure}
\includegraphics[width=0.44\textwidth]{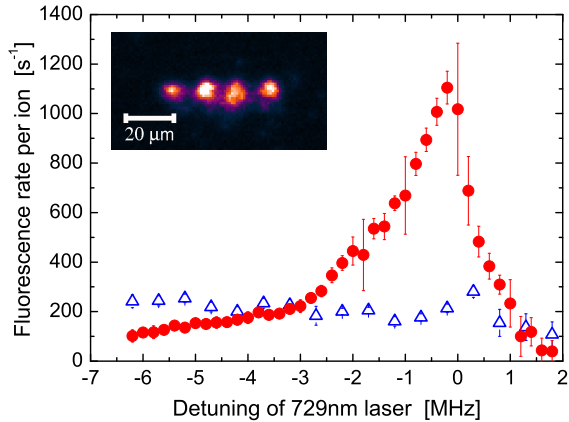}
\caption{\label{fig:covscounter}(Color online.) Plot of detected fluorescence rate from four ions as a function of the 729~nm laser detuning for co-propagating (red, filled circles) and counter-propagating (blue, open triangles) cooling and assisting laser beams. The inset shows a \textrm{CCD} image of the four ions.}
\end{figure}

In figure~\ref{fig:covscounter}, the detected fluorescence rate from a string of four ions is presented as a function of the 729~nm cooling laser detuning.  Each plotted point is an average of 20 consecutive measurements, with the error bars representing the standard deviation of these.  Each individual measurement is preceded by pre-cooling on the dipole-allowed transition and the fluorescence level is measured over 200~ms after these pre-cooling beams are switched off.  For all data points the 729~nm laser beam propagates along the axis of the trap, with the assisting laser beam being either co- or counter-propagating. For the counter-propagating case, the fluorescence level does not change significantly with the detuning, indicating that there is little change in the velocity distribution of the ions. In the co-propagating scenario, however, the fluorescence signal shows a pronounced resonant structure indicating a strong mechanical effect of the light fields. This striking difference, which is not pronounced in resolved sideband cooling, can be explained by the fact that the two-photon momentum kick received by the ion when excited to the P$_{3/2}$ level is about 13 times smaller in the counter-propagating situation than in the case of co-propagating beams. At the largest negative detunings the fluorescence rate for co-propagating beams is clearly lower than for counter-propagating beams, indicating that the ions are so efficiently cooled that they Doppler tune out of resonance.  
%%%%%
%%%%%

\begin{figure}
\includegraphics[width=0.47\textwidth]{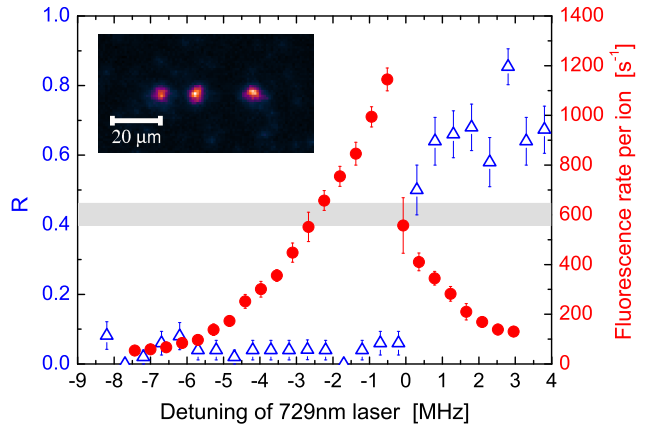}
\caption{\label{fig:jumprates}(Color online.) Plot of the fraction of times $R$ that a non-fluorescing ion in an ion string is seen to switch position as a function of 729~nm laser detuning (blue, open triangles), along with the corresponding fluorescence rate (red, filled circles).  The average jump rate without the 729~nm laser is also shown (shaded horizontal bar).  The inset shows a \textrm{CCD} image of the four ion string, with the gap indicating the position of the non-fluorescing ion.}
\end{figure}

Another way to characterize the cooling effect is to monitor the change in position (jumping) of a single sympathetically-cooled ion in an ion string over time. Only if the kinetic energy of the ions is high enough to overcome the rearrangement potential barrier, typically corresponding to $\sim$1~kelvin, will a jump take place.  In figure~\ref{fig:jumprates}, the fraction of times $R$ that a single non-fluorescing, sympathetically-cooled ion in a four ion string is observed to change postition during a 200~ms period of cooling is presented as a function of the detuning of the 729~nm laser.  The corresponding integrated fluorescence is also plotted.  Clearly there is a dramatic change in the parameter $R$ around zero detuning, which indicates that cooling is indeed taking place at all negative detunings and heating at positive detunings.  This conclusion is further supported by comparison with the value of $R$ when no cooling beams are present.  The non-zero value of $R$ in this case is mainly due to trap-induced heating~\cite{Deslauriers:2004}.  Since the resonance of the 729~nm cooling transition is shifted from that of the unperturbed transition by the presence of the 854~nm assisting laser, we have defined the zero detuning in figures~\ref{fig:covscounter} and~\ref{fig:jumprates} to be the frequency at which $R$ starts to increase steeply.   
    
Based on the observed fluorescence rates and the efficiencies of the photon-detecting devices, the maximum achieved scattering forces in the co- and counter-propagating configurations are estimated to be $4.2 \times 10^{-21}$~N and $3.3 \times 10^{-22}$~N respectively. While the latter is nearly two orders of magnitude smaller than for standard Doppler cooling using the S$_{1/2}$$\longrightarrow$P$_{1/2}$ dipole-allowed transition, the co-propagating geometry gives a force that is only about a factor of 5 lower. The widths of the resonances in the fluorescence are limited by several effects, but assuming Doppler broadening due to the finite temperature of the ions to be dominant a conservative estimate of the ion temperature of a few~mK is obtained.
%%%%%
%%%%%

With the 729~nm cooling laser propagating at an angle of 45$^\circ$ to the axis of the trap, both axial and radial motions of the ions have been cooled. Using such a configuration, single ions have remained laser-cooled for periods of up to a few minutes.  This time seems to be limited by a non-sufficient velocity capture range with the present power level of the 729~nm laser to recool ions that have collided with background atoms or molecules.

\begin{figure}
\includegraphics[width=0.45\textwidth]{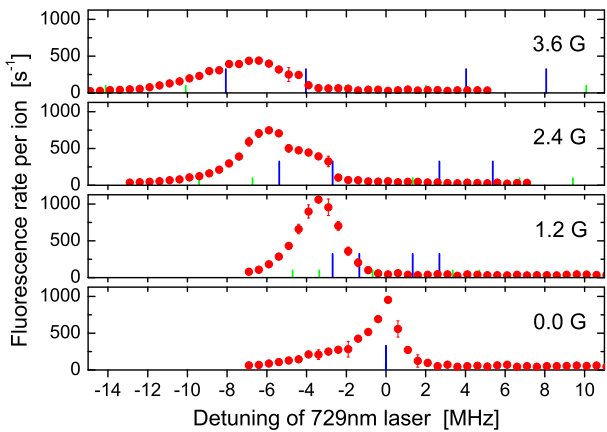}
\caption{\label{fig:magfield}(Color online.) Detected fluorescence as a function of 729~nm laser detuning for different magnetic field strengths.  The vertical blue bars indicate the position of the four $\Delta m_j = \pm 1$ Zeeman components that the laser couples to.}
\end{figure}

A simple way to overcome this problem would be to increase the velocity capture range by increasing the power of the 729~nm laser. At this wavelength, narrow-bandwidth lasers with output powers of several watts have been demonstrated~\cite{Muller:2006}.  A complementary path to enlarge the velocity capture range is to exploit the fact that the S$_{1/2}$$\longrightarrow$D$_{5/2}$ transition actually consists of 10 Zeeman components, which can be split by applying a magnetic field. For a negative detuning of the 729~nm laser with respect to all the Zeeman-split transitions, a cooling force with a maximum value at different velocities will exist for each of the Zeeman lines.  In figure~\ref{fig:magfield}, a series of fluorescence spectra is presented for cooling experiments in the axial cooling geometry with various magnetic field strengths applied parallel to the trap axis.  In this geometry only the four $\Delta m_j = \pm 1$ lines are active~\cite{James:1998}.  As the magnetic field is increased from zero to 1.2 gauss, a significant broadening is observed without any drop in maximum fluorescence.  At higher magnetic fields the Zeeman splitting of the lines becomes comparable to the width of the power-broadened forbidden transition, leading to optical pumping into the more weakly-coupled Zeeman substate.  This results in a lower scattering rate that is visible in the upper two panels of figure~\ref{fig:magfield}.

%%%%%
%%%%%
%%%%%%%%%%%%%%%%%%%%

The presented Doppler cooling scheme could be made slightly simpler by exciting the 733~nm S$_{1/2}$$\longrightarrow$D$_{3/2}$ transition instead of the S$_{1/2}$$\longrightarrow$D$_{5/2}$ transition. In this case the 866~nm laser acts as the assisting laser without need for any repumper laser.  In quantum logic (metrology) experiments where S$_{1/2}$ and D$_{5/2}$ substates are the qubit (relevant) states this scheme would be favorable, since the two cooling lasers can also be used in projection measurements~\cite{Roos:1999,Roos:2006}.

Independent of which transition is used, the benefits of the presented Doppler cooling scheme are manifold: First, it reduces the number of lasers and laser beams needed to eventually obtain ground state cooling.  Only changes to the intensities and frequencies of the two cooling lasers are needed to meet the conditions for resolved-sideband cooling previously demonstrated on the S$_{1/2}$$\longrightarrow $D$_{5/2}$ quadrupole transition~\cite{Roos:1999}. Second, light scattered from the ions is at violet wavelengths that are very different from the red and infra-red excitation wavelengths.  This is equivalent to the situation in two-photon microscopy~\cite{Denk:1990}, and leads to an essentially zero spurious scattering background level.  This feature can be extremely valuable for fast readout of qubits in ion trap quantum information processing~\cite{Leibfried:2005,Haffner:2005}. Third, the requirement that the 729~nm or 733~nm laser have a high intensity and consequently small focus is very compatible with modern micro-traps, in which tight laser focuses are in any case required. These traps have typical electrode spacings of $\sim$100~$\mu$m~\cite{Stick:2005}, and the presented cooling scheme benefits from the excellent overlap between the trapping volume and the laser beams that can be achieved.  Fourth, only lasers in the red and infra-red part of the spectrum, where narrow-bandwidth high power lasers are widely available, are applied. Such technological aspects may play an important role when deciding on a specific ion species for large scale quantum information processing~\cite{Kielpinski:2002}.  Finally, in connection with sympathetic cooling of complex molecular ions, the problem of photofragmentation might in some cases be reduced by using red light instead of violet light~\cite{Kato:2004}.   

Although in the present investigations we have focused on the ${}^{40}$Ca$^+$ ion, the cooling scheme can also be applied to the other isotopes of calcium as well as other ions with a similar atomic level structure, such as Sr$^+$, Ba$^+$, and Hg$^+$~\cite{James:1998}.  For sympathetic cooling of complex molecules, Ba$^+$ is a particularly interesting ion since the S$_{1/2}$$\longrightarrow$D$_{3/2}$ transition is at 2051~nm --- a wavelength at which large molecules often do not absorb and at which powerful commercial fiber lasers exist.
  
In conclusion, a Doppler cooling scheme relying on excitation of the S$_{1/2}$$\longrightarrow$D$_{5/2}$ electric quadrupole transition in the $^{40}$Ca$^+$ ion has been demonstrated. The cooling scheme, which is readily extendable to other alkali earth ion species, contains several features that are interesting for ion trap based quantum information processing and cold complex molecular ion research.  
%%%%%%%%%%%%%%%%%%%%

\begin{acknowledgments}
This work is financially supported by the Carlsberg Foundation and by the EU under contract IST-517675-MICROTRAP.  M.D.~gratefully acknowledges the hospitality of the Laboratoire PIIM-CIML at the Universit\'{e} de Provence under a visiting professor grant.
\end{acknowledgments}

\bibliography{forbidden_cooling_paper}

\begin{thebibliography}{34}
\expandafter\ifx\csname natexlab\endcsname\relax\def\natexlab#1{#1}\fi
\expandafter\ifx\csname bibnamefont\endcsname\relax
  \def\bibnamefont#1{#1}\fi
\expandafter\ifx\csname bibfnamefont\endcsname\relax
  \def\bibfnamefont#1{#1}\fi
\expandafter\ifx\csname citenamefont\endcsname\relax
  \def\citenamefont#1{#1}\fi
\expandafter\ifx\csname url\endcsname\relax
  \def\url#1{\texttt{#1}}\fi
\expandafter\ifx\csname urlprefix\endcsname\relax\def\urlprefix{URL }\fi
\providecommand{\bibinfo}[2]{#2}
\providecommand{\eprint}[2][]{\url{#2}}

\bibitem[{\citenamefont{Adams and Riis}(1997)}]{Adams:1997}
\bibinfo{author}{\bibfnamefont{C.}~\bibnamefont{Adams}} \bibnamefont{and}
  \bibinfo{author}{\bibfnamefont{E.}~\bibnamefont{Riis}},
  \bibinfo{journal}{Prog. Quant. Electr.} \textbf{\bibinfo{volume}{21}},
  \bibinfo{pages}{1} (\bibinfo{year}{1997}).

\bibitem[{\citenamefont{Leibfried et~al.}(2003)\citenamefont{Leibfried, Blatt,
  Monroe, and Wineland}}]{Leibfried:2003}
\bibinfo{author}{\bibfnamefont{D.}~\bibnamefont{Leibfried}},
  \bibinfo{author}{\bibfnamefont{R.}~\bibnamefont{Blatt}},
  \bibinfo{author}{\bibfnamefont{C.}~\bibnamefont{Monroe}}, \bibnamefont{and}
  \bibinfo{author}{\bibfnamefont{D.}~\bibnamefont{Wineland}},
  \bibinfo{journal}{Rev. Mod. Phys.} \textbf{\bibinfo{volume}{75}},
  \bibinfo{pages}{281} (\bibinfo{year}{2003}).

\bibitem[{\citenamefont{Prodan et~al.}(1982)\citenamefont{Prodan, Phillips, and
  Metcalf}}]{Prodan:1982}
\bibinfo{author}{\bibfnamefont{J.~V.} \bibnamefont{Prodan}},
  \bibinfo{author}{\bibfnamefont{W.~D.} \bibnamefont{Phillips}},
  \bibnamefont{and} \bibinfo{author}{\bibfnamefont{H.}~\bibnamefont{Metcalf}},
  \bibinfo{journal}{Phys. Rev. Lett.} \textbf{\bibinfo{volume}{49}},
  \bibinfo{pages}{1149} (\bibinfo{year}{1982}).

\bibitem[{\citenamefont{Chu et~al.}(1985)}]{Chu:1985}
\bibinfo{author}{\bibfnamefont{S.}~\bibnamefont{Chu}} \bibnamefont{et~al.},
  \bibinfo{journal}{Phys. Rev. Lett.} \textbf{\bibinfo{volume}{55}},
  \bibinfo{pages}{48} (\bibinfo{year}{1985}).

\bibitem[{\citenamefont{Neuhauser et~al.}(1978)}]{Neuhauser:1978}
\bibinfo{author}{\bibfnamefont{W.}~\bibnamefont{Neuhauser}}
  \bibnamefont{et~al.}, \bibinfo{journal}{Phys. Rev. Lett.}
  \textbf{\bibinfo{volume}{41}}, \bibinfo{pages}{233} (\bibinfo{year}{1978}).

\bibitem[{\citenamefont{Diedrich et~al.}(1989)}]{Diedrich:1989}
\bibinfo{author}{\bibfnamefont{F.}~\bibnamefont{Diedrich}}
  \bibnamefont{et~al.}, \bibinfo{journal}{Phys. Rev. Lett.}
  \textbf{\bibinfo{volume}{62}}, \bibinfo{pages}{403} (\bibinfo{year}{1989}).

\bibitem[{\citenamefont{Monroe et~al.}(1995)}]{Monroe:1995}
\bibinfo{author}{\bibfnamefont{C.}~\bibnamefont{Monroe}} \bibnamefont{et~al.},
  \bibinfo{journal}{Phys. Rev. Lett.} \textbf{\bibinfo{volume}{75}},
  \bibinfo{pages}{4011} (\bibinfo{year}{1995}).

\bibitem[{\citenamefont{Roos et~al.}(2000)}]{Roos:2000}
\bibinfo{author}{\bibfnamefont{C.~F.} \bibnamefont{Roos}} \bibnamefont{et~al.},
  \bibinfo{journal}{Phys. Rev. Lett.} \textbf{\bibinfo{volume}{85}},
  \bibinfo{pages}{5547} (\bibinfo{year}{2000}).

\bibitem[{\citenamefont{Aspect et~al.}(1986)}]{Aspect:1986}
\bibinfo{author}{\bibfnamefont{A.}~\bibnamefont{Aspect}} \bibnamefont{et~al.},
  \bibinfo{journal}{Phys. Rev. Lett.} \textbf{\bibinfo{volume}{57}},
  \bibinfo{pages}{1688} (\bibinfo{year}{1986}).

\bibitem[{\citenamefont{Lett et~al.}(1988)}]{Lett:1988}
\bibinfo{author}{\bibfnamefont{P.~D.} \bibnamefont{Lett}} \bibnamefont{et~al.},
  \bibinfo{journal}{Phys. Rev. Lett.} \textbf{\bibinfo{volume}{61}},
  \bibinfo{pages}{169} (\bibinfo{year}{1988}).

\bibitem[{\citenamefont{Dalibard and Cohen-Tannoudji}(1989)}]{Dalibard:1989}
\bibinfo{author}{\bibfnamefont{J.}~\bibnamefont{Dalibard}} \bibnamefont{and}
  \bibinfo{author}{\bibfnamefont{C.}~\bibnamefont{Cohen-Tannoudji}},
  \bibinfo{journal}{J. Opt. Soc. Am. B} \textbf{\bibinfo{volume}{6}},
  \bibinfo{pages}{2023} (\bibinfo{year}{1989}).

\bibitem[{\citenamefont{Ungar et~al.}(1989)}]{Ungar:1989}
\bibinfo{author}{\bibfnamefont{P.~J.} \bibnamefont{Ungar}}
  \bibnamefont{et~al.}, \bibinfo{journal}{J. Opt. Soc. Am. B}
  \textbf{\bibinfo{volume}{6}}, \bibinfo{pages}{2058} (\bibinfo{year}{1989}).

\bibitem[{\citenamefont{Shang et~al.}(1990)}]{Shang:1990}
\bibinfo{author}{\bibfnamefont{S.-Q.} \bibnamefont{Shang}}
  \bibnamefont{et~al.}, \bibinfo{journal}{Phys. Rev. Lett.}
  \textbf{\bibinfo{volume}{65}}, \bibinfo{pages}{317} (\bibinfo{year}{1990}).

\bibitem[{\citenamefont{Kasevich and Chu}(1992)}]{Kasevich:1992}
\bibinfo{author}{\bibfnamefont{M.}~\bibnamefont{Kasevich}} \bibnamefont{and}
  \bibinfo{author}{\bibfnamefont{S.}~\bibnamefont{Chu}},
  \bibinfo{journal}{Phys. Rev. Lett.} \textbf{\bibinfo{volume}{69}},
  \bibinfo{pages}{1741} (\bibinfo{year}{1992}).

\bibitem[{\citenamefont{Reichel et~al.}(1995)}]{Reichel:1995}
\bibinfo{author}{\bibfnamefont{J.}~\bibnamefont{Reichel}} \bibnamefont{et~al.},
  \bibinfo{journal}{Phys. Rev. Lett.} \textbf{\bibinfo{volume}{75}},
  \bibinfo{pages}{4575} (\bibinfo{year}{1995}).

\bibitem[{\citenamefont{Binnewies et~al.}(2001)}]{Binnewies:2001}
\bibinfo{author}{\bibfnamefont{T.}~\bibnamefont{Binnewies}}
  \bibnamefont{et~al.}, \bibinfo{journal}{Phys. Rev. Lett.}
  \textbf{\bibinfo{volume}{87}}, \bibinfo{pages}{123002}
  (\bibinfo{year}{2001}).

\bibitem[{\citenamefont{Curtis et~al.}(2001)\citenamefont{Curtis, Oates, and
  Hollberg}}]{Curtis:2001}
\bibinfo{author}{\bibfnamefont{E.~A.} \bibnamefont{Curtis}},
  \bibinfo{author}{\bibfnamefont{C.~W.} \bibnamefont{Oates}}, \bibnamefont{and}
  \bibinfo{author}{\bibfnamefont{L.}~\bibnamefont{Hollberg}},
  \bibinfo{journal}{Phys. Rev. A} \textbf{\bibinfo{volume}{64}},
  \bibinfo{pages}{031403} (\bibinfo{year}{2001}).

\bibitem[{\citenamefont{Salomon et~al.}(1990)}]{Salomon:1990}
\bibinfo{author}{\bibfnamefont{C.}~\bibnamefont{Salomon}} \bibnamefont{et~al.},
  \bibinfo{journal}{Europhys. Lett.} \textbf{\bibinfo{volume}{12}},
  \bibinfo{pages}{683} (\bibinfo{year}{1990}).

\bibitem[{\citenamefont{Mukaiyama et~al.}(2003)}]{Mukaiyama:2003}
\bibinfo{author}{\bibfnamefont{T.}~\bibnamefont{Mukaiyama}}
  \bibnamefont{et~al.}, \bibinfo{journal}{Phys. Rev. Lett.}
  \textbf{\bibinfo{volume}{90}}, \bibinfo{pages}{113002}
  (\bibinfo{year}{2003}).

\bibitem[{\citenamefont{Roos et~al.}(1999)}]{Roos:1999}
\bibinfo{author}{\bibfnamefont{C.~F.} \bibnamefont{Roos}} \bibnamefont{et~al.},
  \bibinfo{journal}{Phys. Rev. Lett.} \textbf{\bibinfo{volume}{83}},
  \bibinfo{pages}{4713} (\bibinfo{year}{1999}).

\bibitem[{\citenamefont{Barton et~al.}(2000)}]{Barton:2000}
\bibinfo{author}{\bibfnamefont{P.~A.} \bibnamefont{Barton}}
  \bibnamefont{et~al.}, \bibinfo{journal}{Phys. Rev. A}
  \textbf{\bibinfo{volume}{62}}, \bibinfo{pages}{032503}
  (\bibinfo{year}{2000}).

\bibitem[{\citenamefont{{S{\o}rensen} et~al.}(2006)}]{Sorensen:2006}
\bibinfo{author}{\bibfnamefont{J.~L.} \bibnamefont{{S{\o}rensen}}}
  \bibnamefont{et~al.}, \bibinfo{journal}{New J. Phys.}
  \textbf{\bibinfo{volume}{8}}, \bibinfo{pages}{261} (\bibinfo{year}{2006}).

\bibitem[{\citenamefont{Kj{\ae}rgaard et~al.}(2000)}]{Kjaergaard:2000}
\bibinfo{author}{\bibfnamefont{N.}~\bibnamefont{Kj{\ae}rgaard}}
  \bibnamefont{et~al.}, \bibinfo{journal}{Appl. Phys. B}
  \textbf{\bibinfo{volume}{71}}, \bibinfo{pages}{207} (\bibinfo{year}{2000}).

\bibitem[{\citenamefont{James}(1998)}]{James:1998}
\bibinfo{author}{\bibfnamefont{D.~F.~V.} \bibnamefont{James}},
  \bibinfo{journal}{Appl. Phys. B} \textbf{\bibinfo{volume}{66}},
  \bibinfo{pages}{181} (\bibinfo{year}{1998}).

\bibitem[{\citenamefont{Barwood et~al.}(1998)}]{Barwood:1998}
\bibinfo{author}{\bibfnamefont{G.~P.} \bibnamefont{Barwood}}
  \bibnamefont{et~al.}, \bibinfo{journal}{Opt. Commun.}
  \textbf{\bibinfo{volume}{151}}, \bibinfo{pages}{50} (\bibinfo{year}{1998}).

\bibitem[{\citenamefont{Deslauriers et~al.}(2004)}]{Deslauriers:2004}
\bibinfo{author}{\bibfnamefont{L.}~\bibnamefont{Deslauriers}}
  \bibnamefont{et~al.}, \bibinfo{journal}{Phys. Rev. A}
  \textbf{\bibinfo{volume}{70}}, \bibinfo{pages}{043408}
  (\bibinfo{year}{2004}).

\bibitem[{\citenamefont{M\"{u}ller et~al.}(2006)}]{Muller:2006}
\bibinfo{author}{\bibfnamefont{H.}~\bibnamefont{M\"{u}ller}}
  \bibnamefont{et~al.}, \bibinfo{journal}{Opt. Lett.}
  \textbf{\bibinfo{volume}{31}}, \bibinfo{pages}{202} (\bibinfo{year}{2006}).

\bibitem[{\citenamefont{Roos et~al.}(2006)}]{Roos:2006}
\bibinfo{author}{\bibfnamefont{C.~F.} \bibnamefont{Roos}} \bibnamefont{et~al.},
  \bibinfo{journal}{Nature} \textbf{\bibinfo{volume}{443}},
  \bibinfo{pages}{316} (\bibinfo{year}{2006}).

\bibitem[{\citenamefont{Denk et~al.}(1990)\citenamefont{Denk, Strickler, and
  Webb}}]{Denk:1990}
\bibinfo{author}{\bibfnamefont{W.}~\bibnamefont{Denk}},
  \bibinfo{author}{\bibfnamefont{J.~H.} \bibnamefont{Strickler}},
  \bibnamefont{and} \bibinfo{author}{\bibfnamefont{W.~W.} \bibnamefont{Webb}},
  \bibinfo{journal}{Science} \textbf{\bibinfo{volume}{248}},
  \bibinfo{pages}{73} (\bibinfo{year}{1990}).

\bibitem[{\citenamefont{Leibfried et~al.}(2005)}]{Leibfried:2005}
\bibinfo{author}{\bibfnamefont{D.}~\bibnamefont{Leibfried}}
  \bibnamefont{et~al.}, \bibinfo{journal}{Nature}
  \textbf{\bibinfo{volume}{438}}, \bibinfo{pages}{639} (\bibinfo{year}{2005}).

\bibitem[{\citenamefont{H\"{a}ffner et~al.}(2005)}]{Haffner:2005}
\bibinfo{author}{\bibfnamefont{H.}~\bibnamefont{H\"{a}ffner}}
  \bibnamefont{et~al.}, \bibinfo{journal}{Nature}
  \textbf{\bibinfo{volume}{438}}, \bibinfo{pages}{643} (\bibinfo{year}{2005}).

\bibitem[{\citenamefont{Stick et~al.}(2005)}]{Stick:2005}
\bibinfo{author}{\bibfnamefont{D.}~\bibnamefont{Stick}} \bibnamefont{et~al.},
  \bibinfo{journal}{Nat. Phys.} \textbf{\bibinfo{volume}{2}},
  \bibinfo{pages}{36} (\bibinfo{year}{2005}).

\bibitem[{\citenamefont{Kielpinski et~al.}(2002)\citenamefont{Kielpinski,
  Monroe, and Wineland}}]{Kielpinski:2002}
\bibinfo{author}{\bibfnamefont{D.}~\bibnamefont{Kielpinski}},
  \bibinfo{author}{\bibfnamefont{C.}~\bibnamefont{Monroe}}, \bibnamefont{and}
  \bibinfo{author}{\bibfnamefont{D.~J.} \bibnamefont{Wineland}},
  \bibinfo{journal}{Nature} \textbf{\bibinfo{volume}{417}},
  \bibinfo{pages}{709} (\bibinfo{year}{2002}).

\bibitem[{\citenamefont{Kato and Yamanouchi}(2004)}]{Kato:2004}
\bibinfo{author}{\bibfnamefont{K.}~\bibnamefont{Kato}} \bibnamefont{and}
  \bibinfo{author}{\bibfnamefont{K.}~\bibnamefont{Yamanouchi}},
  \bibinfo{journal}{Chem. Phys. Lett.} \textbf{\bibinfo{volume}{397}},
  \bibinfo{pages}{237} (\bibinfo{year}{2004}).

\end{thebibliography}

\end{document}